\documentclass{iaus}
\usepackage{graphicx}

\title[LBV Eruptions]{Eruptive Outflow Phases of Massive Stars}

\author[Smith]{Nathan Smith$^1$} \affiliation{$^1$Steward Observatory,
  University of Arizona, 933 N.\ Cherry Ave., Tucson, AZ 85721, USA
  \\email: {\tt nathans@as.arizona.edu}}

\pubyear{2010} \volume{272}
\pagerange{xxx--xxx}
  \jname{Active OB Stars: structure, evolution, mass-loss, and critical
  limits}
  \editors{C.\ Neiner, G.\ Wade, G.\ Meynet \& G.\ Peters, eds.}
\begin{document}
\maketitle

\begin{abstract}

  I review recent progress on understanding eruptions of unstable
  massive stars, with particular attention to the diversity of
  observed behavior in extragalatic optical transient sources that are
  generally associated with giant eruptions of luminous blue variables
  (LBVs).  These eruptions are thought to represent key mass loss
  episodes in the lives of massive stars.  I discuss the possibility
  of dormant LBVs and implications for the duration of the greater LBV
  phase and its role in stellar evolution.  These eruptive variables
  show a wide range of peak luminosity, decay time, expansion speeds,
  and progenitor luminosity, and in some cases they have been observed
  to suffer multiple eruptions.  This broadens our view of massive
  star eruptions compared to prototypical sources like Eta Carinae,
  and provides important clues for the nature of the outbursts.  I
  will also review and discuss some implications about the possible
  physical mechanisms involved, although the cause of the eruptions is
  not yet understood.

\end{abstract}

\firstsection 
\section{Introduction and Background}

Almost sixty years have passed since, as a result of attempts to
produce standard candles for cosmology, Hubble \& Sandage (1953)
discovered the class of luminous, blue, irregular variables in M31 and
M33 that we now collectively refer to as luminous blue variables
(LBVs) in any galaxy.  A few key conferences in the late 1980s and
1990s established some paradigms for LBVs and the evolution of massive
stars in general, some of which may be in need of revision.

In the context of this conference on ``active'' OB stars, the LBVs are
perhaps a hideous extreme example of stellar activity.
However, they can be viewed as cases where the effects of rotation,
pulsation, binaries, and perhaps even magnetic fields may have rather
extreme consequences when the a star is near the Eddington limit.  In
that sense, there is hopefully some synergy between LBVs and the
various other types of stars discussed at this meeting.

LBVs exhibit so-called ``microvariability'' in their photometry and
also undergo well-known S-Doradus excursions when the star changes
color at relatively constant bolometric luminosity (although see the
talk by J.\ Groh in these proceedings).  However, they are most
notable and mysterious for their giant eruptions, when the stars are
thought to increase their bolometric luminosity to be above the
classical Eddington limit, during which time they may eject large
amounts of mass --- anywhere form 0.1 -- 10 M$_{\odot}$.  Smith \&
Owocki (2006) have argued that when this is combined with the facts
that LBV eruptions repeat, and that mass-loss rates for O-type stars
are lower than we used to think, that LBVs probably dominate the
shedding of the H envelope in massive single stars.  This may have
significant implications, since LBV eruptions do not necessarily
depend on metallicity.

However, we still have no clear idea what causes the giant eruptions
of LBVs, and we have no good formulation for how the eruptive behavior
scales with initial mass and metallicity, or if it depends on
binarity.  Since the dominant mass-loss machanism in stellar evolution
is so poorly understood, we cannot have very much faith in the
predictions of the fates of massive stars in stellar evolution models,
or how this scales with metallicity.

We do, however, know that giant LBV eruptions certainly occur because
we observe them, and advances can be made in constraining their
properties.  Giant LBV eruptions are bright and can be seen in other
galaxies.  They are detected by accident in systematic supernova
searches that are conducted --- like Hubble \& Sandage's early work
--- in the pursuit of standard candles for cosmology, and so they are
sometimes called ``SN impostors''.  Several dozen SN impostor giant
eruptions have now been seen in nearby galaxies, but LSST will vastly
increase the number of these transients.  Hopefully this will allow us
to improve our knowledge of the statistics of LBVs.  For now, we must
be content with studying the few examples we have and gleaning as many
clues about their physics as we can.  In this paper, I briefly review
some of the observed properties of LBV stars, and I emphasize some new
results including the distribution of observed properties in giant LBV
eruptions and their connection to Type~IIn supernovae.  Much of what I
discussed in my talk at IAU Symposium 272 is presented in more detail
in two recent papers (Smith et al.\ 2010a; Smith \& Frew 2010), and
the reader is reffered to these for more information.

\section{Lifetime of the LBV Phase, and Ducks that Don't Quack}

``If it looks like a duck, and quacks like a duck, we have at least to
consider the possibility that we have a small aquatic bird of the
family anatidae on our hands.''  

...Douglas Adams

\smallskip \smallskip

\begin{figure}[b]
\begin{center}
 \includegraphics[width=5.9in]{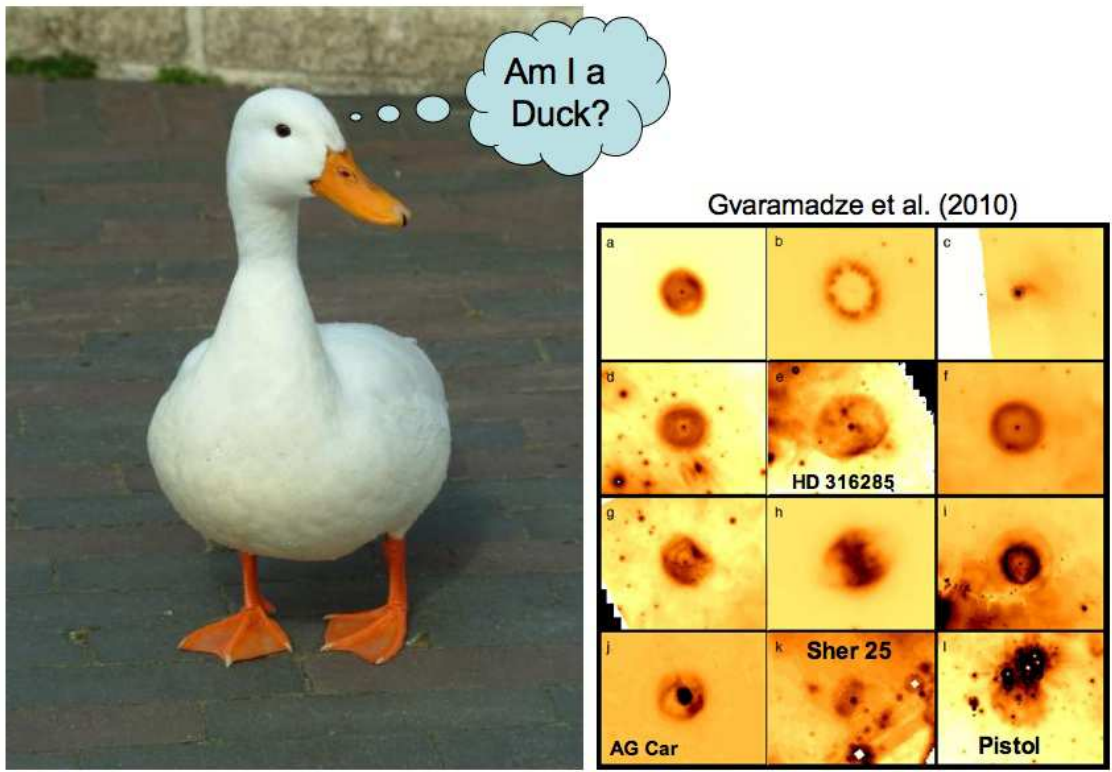} 
 \caption{The reader can deduce that the object on the left is
   obviously a Duck, even though this conference proceedings volume is
   not accompanied by an audio CD containing a recording of it
   quacking.  On a related note, the panel at right shows several hot
   massive stars surrounded by LBV-like dust shells detected in the
   mid-IR (Gvaramadze et al.\ 2010).}
   \label{fig1}
\end{center}
\end{figure}

This is a slightly different formulation of the more familiar ``If it
looks like a duck...'' phrase, which was often used in connection with
LBV eruptions in the 1980s and 1990s (e.g., Conti 1995, 1997; Bohannan
1997), suggesting that you can't really be sure that a duck is a duck
unless you hear it quack.  The point was that although pretty much
everything in the upper left part of the HR diagram is luminous, blue,
and at least somewhat variable if you look closely enough, the
classification ``LBV'' was to be reserved for a specific class of
stars that are observed to undergo more violent eruptions (i.e. they
``quack'' rather loudly), and that this therefore indicated some
particular inherent instability in the star, which is not present in
all supergiants.  In the same breath, however, it was sometimes
admitted that a star which had erupted in the past (or will erupt in
the near future) might not necessarily be exhibiting signs of that
instability {\it right now}.  This problem is illustrated in
Figure~\ref{fig1}.

LBVs are extremely rare --- there are only a handful known in the
Milky Way or in any nearby galaxy.  However, there is a larger number
of stars that closely resemble LBVs in their observed spectral
properties and location on the HR diagram.  Some of these have massive
circumstellar shells that resemble LBV shells, as in several recent
examples detected by Spitzer (Gvaramdze et al.\ 2010; Wachter et al.\
2010).  These are typically called ``LBV candidates'' until they are
actually seen to have an eruption.  The Ofpe/WN9 stars are a good
example of a class of stars which resemble LBVs and often have
circumstellar shells; there are documented examples of confirmed LBVs
that are Ofpe/WN9 stars in their quiescent hot states (e.g., R127, AG
Car).  If Ofpe/WN9 stars are really dormant LBVs, it would imply that
LBVs may go through relatively long periods of time when they are not
erupting.  In other words, they may have extended ``dormant'' phases
in between major eruptions, like volcanoes or geysers.

Another sign of long dormant phases of LBVs is illustrated by the
example of P~Cygni, which is the nearest and the first LBV.  It
underwent a giant LBV eruption in 1600 AD, with a second eruption 55
years later.  After being faint for another 50 years, it brightened at
the beginning of the 18th century.  Since then, however, {\it P~Cygni
  hasn't done much of anything to suggest that it is an unstable
  star}.  Had we started observing it around 1700 AD, instead of 1600
AD, we would never know that it is an LBV.  We would see a relatively
tame blue supergiant with a strong wind and a shell nebula, and we
would simply call it an LBV candidate. (Note that P~Cygni is not as
hot as an Ofpe/WN9 star, suggesting that there may be many other blue
supergiants that are dormant LBVs as well.)  Similarly, many stars
discovered in the Galactic center resemble LBVs in their luminosity
and spectral properties, but because they are only seen in the IR, we
do not have the benfit of many decades or centuries of continuous
photometric monitoring, so we have not necessarily observed eruptive
behavior in all the Galactic center LBVs.
 
The rarity of LBV has led to important conjectures about their
evolutionary phase and the lifetime of the LBV phase itself.  Many
authors have discussed this, but the argument usually goes something
like this, as discussed in the review by Bohannan (1997): There were 5
LBVs and 115 WR stars known in the LMC at that time.  If we assume
that all WR stars are descended from LBVs, and that the WR lifetime is
the core-He burning lifetime of about 5$\times$10$^5$ yr, the rarity
of LBVs then suggests that the LBV phase only lasts only a few 10$^4$
yr.  It was widely concluded, therefore, that the LBV phase represents
an extremely brief and fleeting {\it transitional} phase between the
core-H burning main sequence of O-type stars and the core-He burning
phase of WR stars.

This line of reasoning suffers from some fallacies, and the derived
age is probably wrong.  It ignores the possibility of dormant phases
of LBVs, as noted above, and offers no good explanation for the large
number of LBV candidates and other blue supergiants that are
necessarily evolved massive stars as well.  It also ignores the fact
that about 1/3 of stars counted as ``WR stars'' are actually WNH stars
(see Smith \& Conti 2008) and are probably not in core-He burning yet.

Returning to the analogy with volcanoes, one could reproduce a similar
fallacy: there are typically something like 1 or 2 major volcanic
eruptions on Earth each year (where ``major'' means more than 0.1
km$^3$ of tephra).  Some of us experienced the unfortunate
consequences of this for international travel earlier this year.  One
could then say that since there are several thousand major mountains
on Earth, each of which has an average geological age of around 10$^8$
yr, that the lifetime of a typical volcano is only a few 10$^4$ yr.
This is, of course, a severe underestimate for the lifetime of a
volcano because volcanoes spend most of their time in dormant phases.
We know this because mountains with a crater or with evidence for a
history of eruptions are counted as real volcanoes, and may erupt
again in the future.  Similarly, one could take inventory of the
number of ducks quaking at any instant and vastly underestimate the
true number of aquatic birds of the family anatidae.

Deriving the correct lifetime for LBVs depends on the ``duty cycle''
of the unstable LBV phase.  In other words, we need to know what
fraction of the time an LBV might be dormant by our observational
standards, and correct for that.  We have no theoretical prediction of
this time, since there is no theoretical prediction of LBVs.  There
is, however, an expectation that LBVs recover from major eruptions and
go through a relatively quiescent period where they re-establish
thermal equilibrium.  Both P~Cygni and $\eta$ Car have mulitple shell
nebulae that suggest time periods of order 10$^3$ yr in between major
eruptions.  Moreover, Massey et al.\ (2007) counted only 6 LBVs in M31
and M33, but they counted over 100 LBV candidates.  This suggests a
factor of 10-20 more LBVs than are counted by active LBVs at any time,
implying a duty cycle of 5--10\% for the manifestation of LBV
instability during the greater evolutionary phase in which we find
LBVs.  If we re-do the calculation above (now including the fact that
1/3 of WR stars are WNH), then we find that the lifetime over which a
massive star could be an LBV is more like (2--5)$\times$10$^5$ yr.

This paints a very different picture for the evolutionary state of
LBVs, where they spend a substantial fraction (or all) of their
core-He burning lifetime as an LBV (or candidate LBV), punctuated by
intermittent episodes of eruptive instability.  If some of these LBVs
make it all the way to core collapse before shedding their H
envelopes, it may explain the observed connection between LBVs and
Type~IIn supernovae (Smith et al.\ 2007, 2008, 2010a, 2010c; Gal-Yam
\& Leonard 2009; etc.).

Of course, the comments above are predicated on the notion that all WR
stars are descended from LBVs, allowing us to calculate the LBV
lifetime by comparison to the assumed WR lifetime.  This hypothesis
may be wrong if, for example, a substantial fraction of WR stars have
shed their H envelopes via Roche lobe overflow in binary systems (see,
e.g., Smith et al. 2010c for implications from Type Ibc supernovae).
In that case, the fraction of LBVs+candidates to WR stars depends on
both the relative lifetimes and the fraction of massive stars in close
binaries.  One gets the impression that our paradigms of massive star
evolution need to be taken back to the drawing board.

\begin{figure}[b]
\begin{center}
 \includegraphics[width=6.0in]{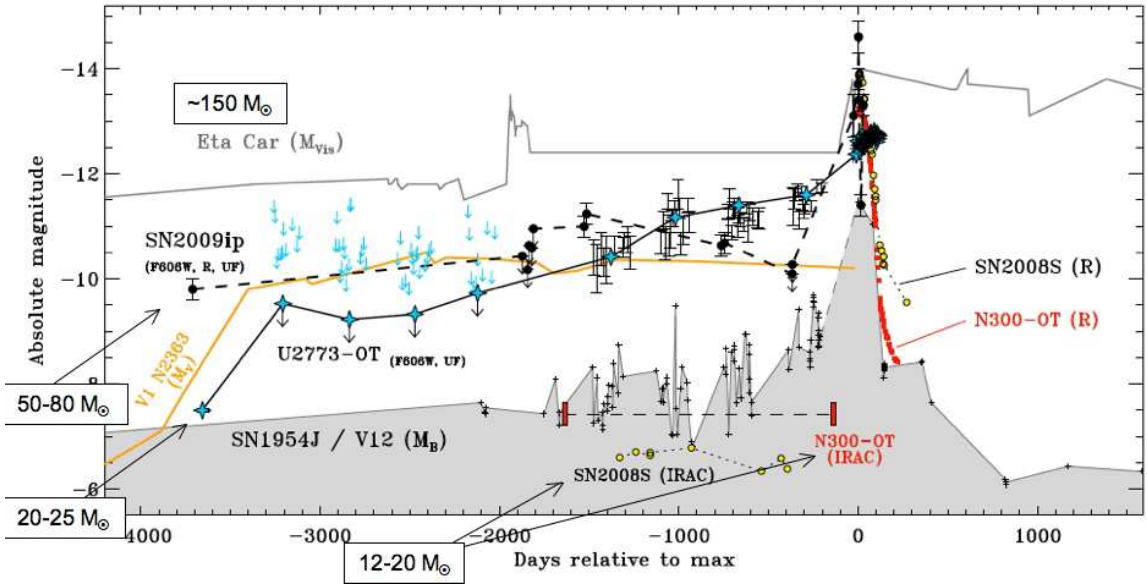} 
 \caption{Light curves of various LBV-like transients (from Smith et
   al.\ 2010a), with a few cases where approximate initial masses for
   the progenitor stars have been estimated.  References for
   individual sources of photometry can be found in that paper.}
   \label{fig2}
\end{center}
\end{figure}

\begin{figure}[b]
\begin{center}
 \includegraphics[width=3.9in]{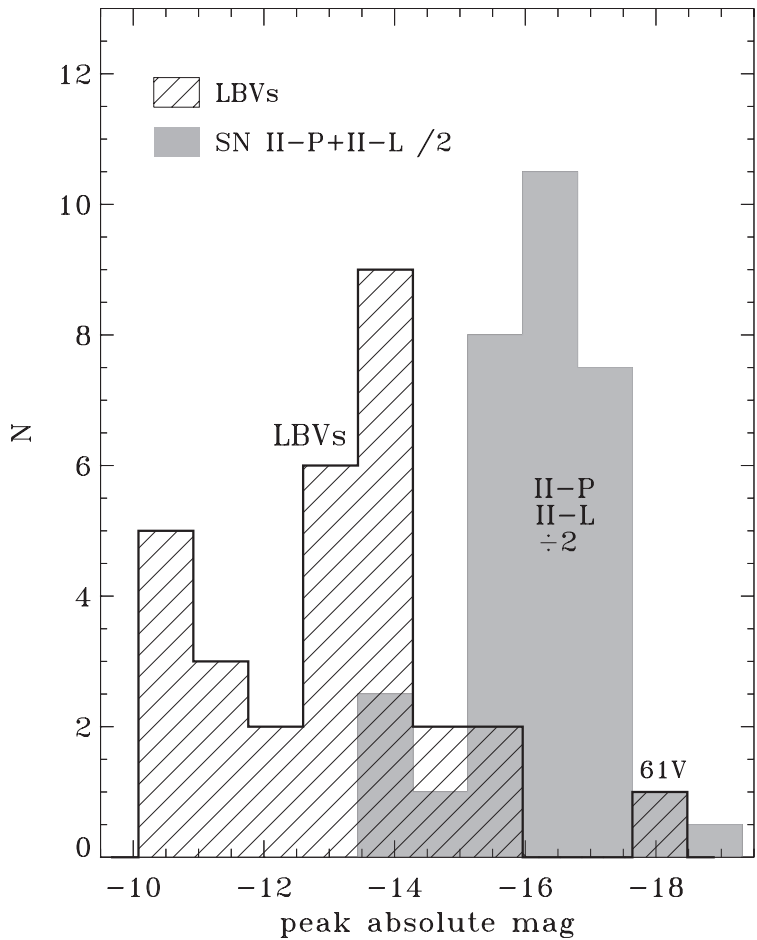} 
 \caption{A histogram of the distribution of peak absolute magnitudes
   for LBV-like eruptions (hatched) compared to normal Type II-P and
   II-L supernovae (gray; numbers divided by 2) from the Berkeley SN
   search (figure from Smith et al.\ 2010a; see that paper for
   details).  The LBV farthest to the right is SN~1961V, which Smith
   et al.\ (2010a) have argued is not really an LBV, but rather, a
   true core-collapse SN IIn.}
   \label{fig3}
\end{center}
\end{figure}

\begin{figure}[b]
\begin{center}
 \includegraphics[width=4.1in]{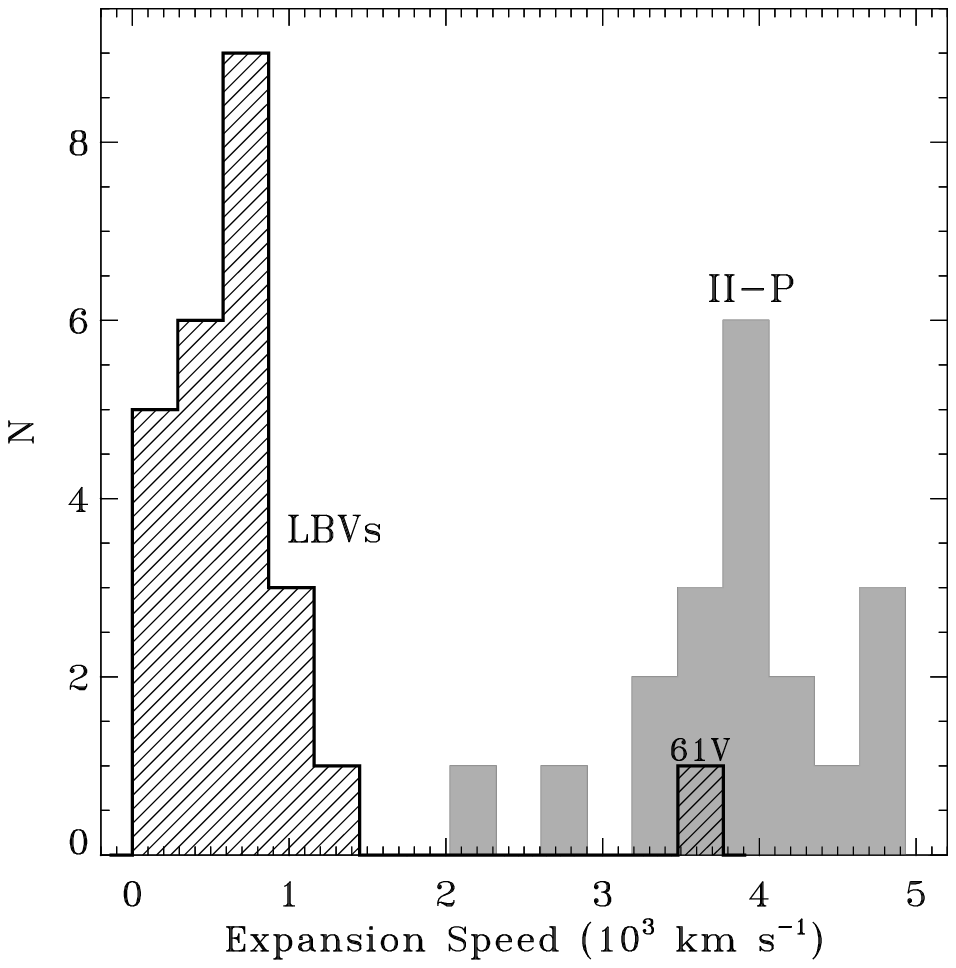} 
 \caption{A histogram of the expansion speeds for LBV-like eruptions
   (hatched) compared to SNe II-P (figure from Smith et al.\ 2010a;
   see that paper for details).}
   \label{fig4}
\end{center}
\end{figure}

\section{A Diverse Range of Observed Properties}

LBVs are by definition associated with the most luminous and most
massive stars in any galaxy, but their initial mass range is actually
rather wide.  LBVs are thought to arise from stars with initial masses
ranging from 20 or 25 M$_{\odot}$ up to the most massive stars known
(Smith, Vink, \& de Koter 2004).  (The lower-luminosity LBVs with
initial masses of 20 or 25 $M_{\odot}$ up to about 40 $_{\odot}$ are
thought to reach their unstable state only after they have been
through substantial mass loss in a previous RSG phase, thereby
increasing their L/M ratio.)  This mass range is perhaps a result of
how we identify them: we define the LBV variability as a brightening
at visual wavelengths that corresponds roughly to the star's
bolometric correction (e.g., Humphreys \& Davidson 1994).  The
S~Doradus instability strip is slanted on the HR diagram, so that more
luminous LBVs are hotter in their quiescent state.  As a result, these
hotter and more luminous LBVs have a larger bolometric correction, and
consequently, brighten more at visual wavelengths when they undergo an
S~Dor eruption.\footnote{Note that it was originally hoped that
  calibrating this would allow LBVs to be used as standard candles.}
These are the classical LBVs.  LBVs at the bottom end of the initial
mass range have cooler quiescent temperatures and, consequently,
smaller bolometric correction and less pronounced brightenings in a
normal S~Dor stage.  In fact, if we were to extrapolate the S~Dor
instability strip to lower luminosities and cooler temperatures, it
would cross the temperture for cool 7000--8000 K eruptive states of
LBVs at luminosities that correspond to initial masses of $\sim$20
$M_{\odot}$.  In other words, whatever instability causes the LBV
phase might manifest itself somewhat differently below 20--25
$M_{\odot}$, and we might not recognize these stars as LBVs because of
how we define the observed LBV variability.


In fact, recent studies of extragalactic transient sources have
revealed some transients that closely resemble LBV giant eruptions,
but which --- unexpectedly --- seem to have progenitor stars of lower
masses around 10--20 $M_{\odot}$ or even lower.  These transients and
other LBVs are reviewed recently by Smith et al.\ (2010a; see also
Smith et al.\ 2009, 2010b; Prieto et al.\ 2008, 2009; Thompson et al.\
2009; Gogarten et al\ 2009).  The true nature of these sources is
still debated, however; it is not clear if they are manifestations of
LBV-like instability extending to stars with lower-initial masses, or
if they are something altogether different originating from
intermediate-mass stars.  The light curves for some LBV-like
transients are shown in Figure~\ref{fig2} (from Smith et al.\ 2010b),
concentrating on some sources that show detections of their progenitor
stars before a giant LBV-like eruption.  This is meant to demonstrate
the range of initial luminosities and masses for stars that undergo
giant LBV-like eruptions.  Some of the stars even show precursor
variability before the eruption begins, like SN~2009ip and UGC2773-OT
(Smith et al.\ 2010b).

Smith et al.\ (2010a) has also discussed the diversity in the observed
properties of the eruptions themselves.  Figures~\ref{fig3} and
\ref{fig4} show histograms of the distributions of peak absolute
visual magnitude (a combination of $V$ and $R$ magnitudes) and the
distribution of expansion speeds (measured from H$\alpha$).  LBV-like
eruptions span a range in absolute peak magnitude from around --10 to
-16 mag, peaking at --14 mag.  The luminous end of the distribution
overlaps with the faintest core-collapse SNe, but one can usually
distinguish the two based on spectra (see Smith et al.\ 2009, 2010b).
The low-luminosity end of the distribution of LBV-like eruptions is
muddy; there may be a mix of LBV giant eruptions and S Doradus
outbursts (see Smith et al.\ 2010a for more details).  One problematic
case is the prototypical SN impostor SN~1961V, which is much brighter
than any other LBV eruption.  SN~1961V also stands out in its observed
expansion speed (Figure~\ref{fig4}) which is much faster than other
LBVs and more in line with core-collapse SNe.  Based on these points
and other information, Smith et al.\ (2010a) have argued that SN~1961V
was in fact not an LBV giant eruption, but a true core-collapse
SN~IIn.  The other LBVs have expansion speeds that range from around
100 to 1000 km s$^{-1}$, much slower than speeds for core-collapse
SNe, indicating less energetic explosions.

\begin{figure}[b]
\begin{center}
 \includegraphics[width=6.0in]{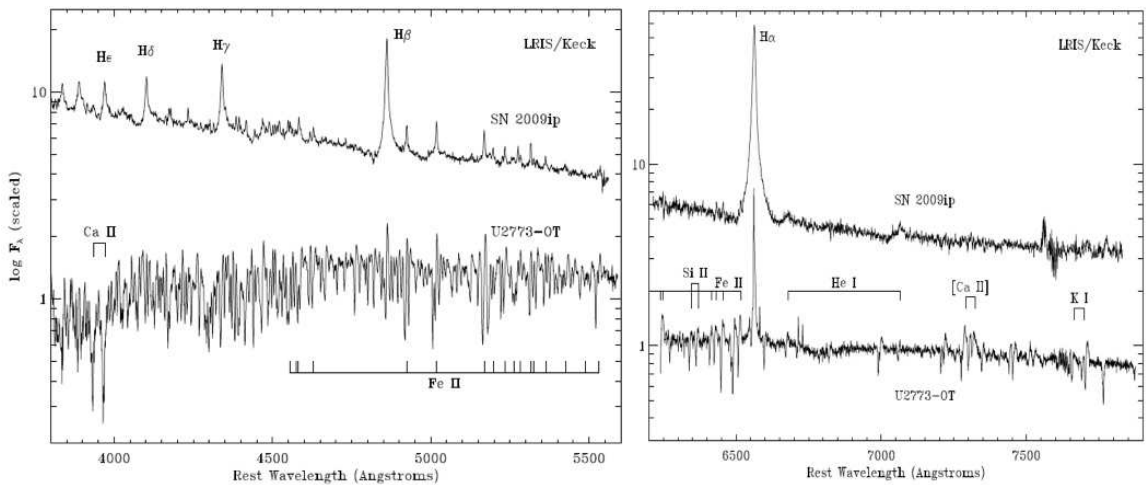} 
 \caption{Visual-wavelength spectra of two recent transients,
   demonstrating the range of spectral properties in LBV outbursts.
   SN~2009ip is an example of a hotter LBV, with blue continuum and
   strong and relatively broad Balmer emission lines, plus evidence
   for a blast wave.  UGC~2773-OT is more characteristic of cooler LBV
   wind-dominated spectra in their F supergiant state.  Both are from
   Smith et al.\ (2010b).}
   \label{fig5}
\end{center}
\end{figure}

\section{Some Detailed Examples}

There has been a recent increase in studies of extragalactic
transients that seem analogous to LBV giant eruptions, perhaps due in
part to the increased community-wide interest in transient sources of
all types, and perhaps also because a substantial fraction of the SN
community seems to finally be getting bored of Type~Ia SN cosmology.
Whatever the reason, extragalacitc LBV-like eruptions are receiving
more attention and we have more examples of them, with the result that
the increased number do not support some long-help paradigms about
LBVs.

\begin{figure}[b]
\begin{center}
 \includegraphics[width=6.0in]{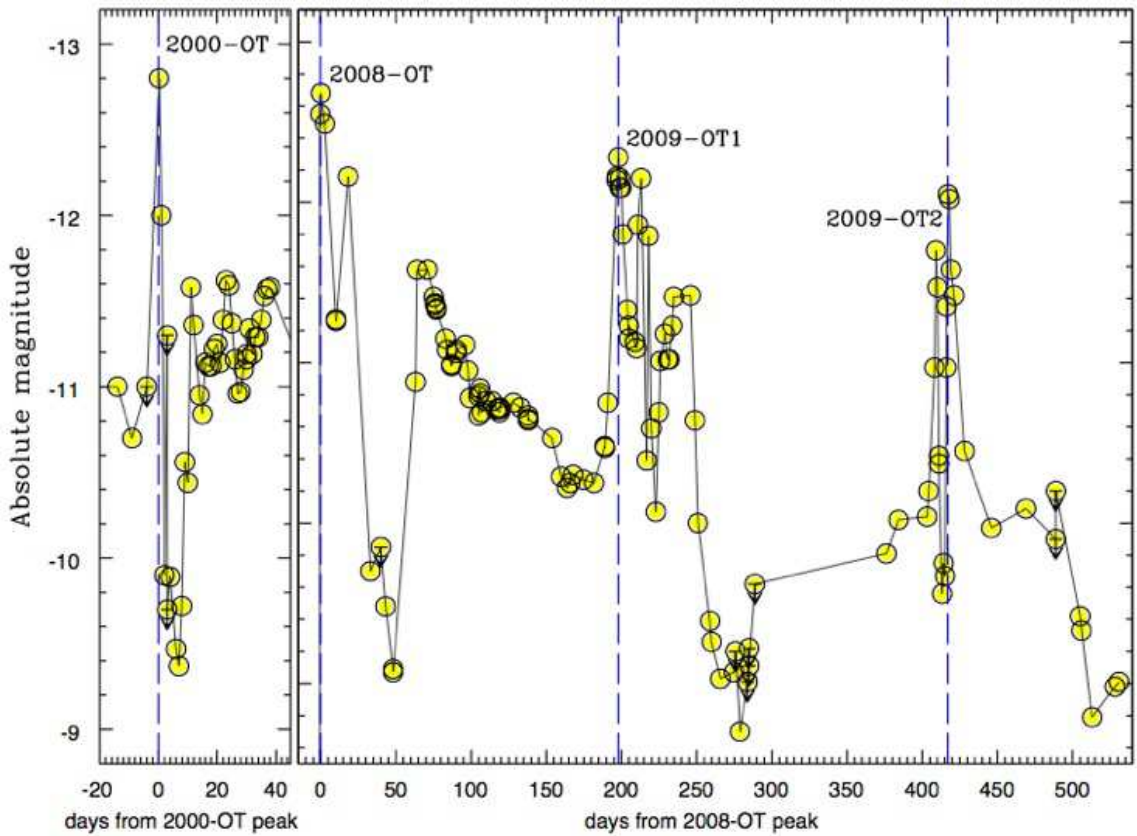} 
 \caption{The $R$-band light curve of the multiple eruptions of
   SN~2000ch (in 2000, and then again in 2008-2009), from Pastorello
   et al.\ (2010).}
   \label{fig6}
\end{center}
\end{figure}

In particular, LBV eruptions were thought to always have cool
$\sim$8000 K pseudo photospheres (Humphreys \& Davidson 1994), but
this is apparently wrong.  Some do indeed exhibit apparent
temperatures in this range and have F-supergiant like spectra; the
recent transient UGC2773-OT is a good example, and its spectrum is
shown in Figure~\ref{fig5}.  However, several LBV giant eruptions
exhibit hotter temperatures with smooth blue continua and strong
Balmer emission lines.  Some even show evidence for fast blast waves
of 5,000 km s$^{-1}$ out in front of the bulk of ejecta moving at
around 600 km s$^{-1}$.  SN~2009ip is an example of this
(Figure~\ref{fig5}; Smith et al.\ 2010b), as is the more familiar case
of $\eta$ Carinae (Smith 2008).  The spectral diversity of these LBV
giant eruptions is discussed in more detail by Smith et al.\ (2010a).

An exciting recent development is that some LBVs exhibit multiple
brief eruptions, partly as a consequence of continued monitoring.  It
was already known that both $\eta$ Car and P Cygni had secondary
eruptions about 50 yr after their initial giant eruptions (Humphreys
\& Davidson 1994; Humphreys et al.\ 1999).  However, a recent study by
Smith \& Frew (2010) shows an even more complicated situation for
$\eta$~Car, with two brief precursor eruptions in 1838 and 1843 that
preceded the peak of the Great Eruption in 1844, and which appear to
have occurred near times of periastron.  Moreover, the LBV-like
eruption SN~2000ch (see Wgner et al.\ 2004) was later discovered to
have multiple subsequent eruptions in 2008 and 2009 (Pastorello et
al.\ 2010).  Very recently, Drake et al.\ (2010) reported the
discovery of another subsequent eruption of SN~2009ip, which was
discussed above.  In all cases, the repeated eruptions appear to be
very brief (few to 10$^2$ days), and not the multi-year affairs as
seen in more conventional LBV eruptions.  The physical cause of these
repeated brief outbursts is not yet known, but Smith et al.\ (2010a)
and Pastorello et al.\ 92010) have mentioned the possibility of binary
interactions, among other potential causes.  Smith (2010) discusses a
particular way that such a model might work.

Lastly, there is continually mounting evidence that LBV-like eruptions
seem to precede the particular class of supernovae known as Type IIn,
where the narrow (n) lines of H arise when the SN blast wave
encounters extremely dense circumstellar material ejected immediately
before the outburst.  This has been discussed elsewhere by multiple
authors at previous conferences.  The point I would like to emphasize
here is that the range of properties inferred for the precursor
eruptions of SNe IIn seems to roughly match the diversity in
properties exhibited by LBV eruptions themselves (expansion speed,
mass-loss rates, composition), but there are no other known stars with
sufficient mass-loss rates to match SN~IIn progenitors.  The SN
IIn/LBV connection will likely become clearer with more studies of LBV
eruptions and of SNe IIn, and especially cases where LBV progenitor
stars are seen to explode as SNe IIn (e.g., Gal-Yam \& Leonard 2009).

\begin{discussion}



\end{discussion}

\end{document}